\documentclass[useAMS,usenatbib,referee]{mnras}

\usepackage{amsmath}
\usepackage{amssymb}
\usepackage{graphicx}
\usepackage{dcolumn}
\usepackage{bm}
\usepackage{ulem}
\usepackage{color}

\newcommand{\BV}{Brunt-V{\"a}is{\"a}l{\"a}}
\newcommand{\kB}{k_{\rm B}}
\newcommand{\Msun}{\rm M$\odot$}

\title[Quantum ion thermodynamics of white dwarfs] 
{Quantum ion thermodynamics in liquid interiors 
of white dwarfs}

\author[D. A. Baiko and D. G. Yakovlev] 
{D.A. Baiko\thanks{E-mail:baiko.astro@mail.ioffe.ru} and
D.G. Yakovlev\thanks{E-mail:yak.astro@mail.ioffe.ru} \\
Ioffe Institute, Politekhnicheskaya 26, 194021 Saint Petersburg, 
Russian Federation}

\begin{document}

\label{firstpage}
\date{Accepted; Received ; in original form}

\pagerange{\pageref{firstpage}--\pageref{lastpage}} \pubyear{2019}

\maketitle

\begin{abstract}
We present an accurate analytic approximation for the energy of a quantum 
one-component Coulomb liquid of ions in a uniform electron background 
which has been recently calculated from first principles (Baiko 2019). 
The approximation enables us to develop in an analytic form a complete 
thermodynamic description of quantum ions in a practically important 
range of mass densities at temperatures above crystallization. 
We show that ionic quantum effects in liquid cores of white dwarfs (WDs) 
affect heat capacity, cooling, thermal compressibility, pulsation 
frequencies and radii of sufficiently cold WDs, especially
with relatively massive helium and carbon cores.
\end{abstract}


\section{Introduction}
\label{s:introduc}

Thanks to {\it Sloan Digital Sky Survey}, {\it Gaia} and other projects, current progress in 
highly precise observations and theoretical interpretation of WDs is fantastically 
rapid (e.g., \citealt{GB19,BD19,KP19,Pel19}). 
For instance, we can mention studies of cooling WDs  
(e.g., \citealt{TFF19,FV19,Blouin19}) and asteroseismology (e.g., \citealt{CABK19,BK19}). 
The rapid observational progress motivates 
further theoretical studies of dense matter in WD interiors.  

Recently, \citet{B19} (hereafter Paper I) has calculated from first principles the energy of 
a liquid quantum one-component plasma (OCP) of ions immersed in a uniform charge-compensating background
of electrons. This is a suitable model to describe dense plasma of ions in 
degenerate cores of WDs. The author presented his results in a tabular form and illustrated them
by calculating temperature dependence of the ion specific heat at two combinations of density and composition
(using an interpolation of the tabulated data). Also, he pointed out a number of
applications of these results for WDs (as well as for the envelopes of neutron stars).  

In Section \ref{s:fit}, we present an analytic fit to the results of Paper I. 
It allows us to construct thermodynamics of a quantum one-component ion liquid in 
a closed form thereby facilitating usage of the results of Paper I in applications. In Section
\ref{s:applic}, we employ these fitting formulae to study quantitatively the ionic 
quantum effects in liquid cores of WDs and assess their importance. 
We conclude in Section \ref{s:conclude}.

\section{Analytic formulation}
\label{s:fit}

\subsection{Ion plasma parameters}

Thermodynamic state of the OCP of ions immersed in a rigid
electron background can be characterized by two dimensionless 
parameters (e.g., \citealt{HPY2007}),
\begin{equation}
    \Gamma = \frac{Z_i^2 e^2}{\kB Ta}, \quad \eta=\frac{T_{\rm p}}{T \sqrt{3}}.
\label{e:key}    
\end{equation} 
Here, $Z_i$ is the ion charge number, $e$ is the elementary charge,
and $\kB$ is the Boltzmann constant. Furthermore, $T$ is the temperature,
$a = (4 \pi n_i/3)^{-1/3}$ is the ion-sphere radius determined
by the number density of ions $n_i$;  
$T_{\rm p} = (\hbar/\kB) \sqrt{4 \pi n_i Z_i^2 e^2 /M_i}$ 
is the ion plasma temperature and $M_i$ is the ion mass. 

The first quantity, the Coulomb-coupling parameter 
$\Gamma$, measures the ratio of the typical 
potential and kinetic energies of ions. If $\Gamma \ll 1$, the ions are weakly
coupled (constitute Boltzmann gas). In the opposite limit 
of $\Gamma \gg 1$, they form
a strongly coupled Coulomb plasma, liquid or solid. If one disregards
quantum effects in ion motion, melting occurs at 
$\Gamma=\Gamma_{\rm m} \approx 175$ \citep{PC00}.
The melting temperature is 
\begin{equation}
      T_{\rm m}=\frac{Z_i^2e^2}{\kB a \Gamma_{\rm m}}.
      \label{e:Gamma_m}
\end{equation}

The second parameter $\eta$ in equation (\ref{e:key}) characterizes the strength 
of the ion quantum effects. These effects are known to be especially important for
sufficiently light ions at relatively low temperatures and/or high densities ($\eta \gtrsim 1$).
In a quantum system, one has $T \ll T_{\rm p}$. Another familiar quantum-mechanical parameter is 
$r_{\rm s} = a / a_0$, where $a_0 = \hbar^2 /(M_i Z_i^2 e^2)$ is the 
ionic Bohr radius. These quantities are related as
\begin{equation}
\eta=\frac{\Gamma}{\sqrt{r_{\rm s}}} ~.
\label{e:rs}
\end{equation}

\subsection{Analytic fit}
\label{s:fitting}

Energy $E_i$ of a quantum Coulomb plasma of ions with uniform 
incompressible electron background was calculated in Paper I (table 1) 
neglecting ion exchange effects
on a dense grid of values of $\Gamma$ 
($1 \leq \Gamma \leq 175$, 25 grid points) and $r_s$ 
($600 \leq r_s \leq 12000$, 15 points; 25$\times 15=
375$ grid points in total). Here, these results are   
approximated by the following analytic expression,
\begin{equation}
    \frac{E_i - E_0}{N_i \kB T} = \frac{3}{2} - \zeta \Gamma 
     + u_{\rm cl}(\Gamma) + u_{\rm q}(\eta)~. 
\label{fit_en}
\end{equation}
The functions which enter equation (\ref{fit_en}) are described below 
and the fit parameters are given in Table \ref{coeff}.
The relative root mean square error of the fit over all grid 
points is $0.0014$ while the maximum relative error of 0.0043 
takes place at $\Gamma=73.5$ and $r_s=4800$. Such fit accuracy 
seems acceptable for astrophysical applications. 
An explicit formula is much easier to use than tabulated values; it allows 
one to differentiate and integrate analytically over temperature and 
density for constructing 
various thermodynamic quantities of the ion plasma.
  
The quantity on the left-hand side of equation (\ref{fit_en}) was calculated and 
tabulated in Paper I [where it was denoted as 
$(E-E_0)/(NT)$ whereas $\kB$ was set equal to 1], $E_0 = \zeta \Gamma N_i \kB T$ is the 
electrostatic (Madelung) energy of 
an ideal body centered cubic (bcc) Coulomb crystal ($\zeta \approx -0.895929255682$), and 
$N_i$ is the number of ions. 
Furthermore, $u_{\rm cl}(\Gamma)$ represents the fit proposed by \citet{PC00} 
for their `ion-ion' component of the energy of a classic Coulomb plasma, 
\begin{equation}
     u_{\rm cl}(\Gamma) 
     = \Gamma^{3/2} \left[\frac{A_1}{\sqrt{A_2+\Gamma}} 
     + \frac{A_3}{1+\Gamma} \right]  
  +\frac{B_1 \Gamma^2}{B_2 + \Gamma} 
  + \frac{B_3 \Gamma^2}{B_4 + \Gamma^2}~,
\label{pot_ds}
\end{equation} 
with $A_3 = -\sqrt{3}/2 - A_1/\sqrt{A_2}$. 
At $\Gamma > 1$, the fit (\ref{fit_en}) {\it without} the last term 
$u_{\rm q}$ reproduces Monte Carlo (MC) results of 
\citet{DS99} while at $\Gamma < 1$, it interpolates between 
these MC results and a well known exact analytic expression. In the latter case quantum
corrections become insignificant; accordingly, 
equation (\ref{fit_en}) can be used at arbitrarily small $\Gamma$. 

The quantum contribution to the energy is described by the last term
of equation (\ref{fit_en}),
\begin{equation}
   u_{\rm q}(\eta) = 
    \frac{Q_1 \eta^2}{Q_2 + \eta} + \frac{Q_3 \eta^2}{Q_4 + \eta^2}~.
\label{dg_b}
\end{equation}
It turns out to be a function of a single parameter
$\eta=T_{\rm p}/(\sqrt{3}T)$
given by equation (\ref{e:rs}).
This fact seems nontrivial; it has not been expected from the beginning but
the fit procedure indicates that it is so (at least within the fit
accuracy). In the limit $\eta \to 0$, the quantum term 
is forced to reduce to the exact first-order Wigner-Kirkwood correction (e.g., \citealt{LL80}).
This translates to 
the condition $Q_1/Q_2+Q_3/Q_4 = 1/4$ which allows one to 
calculate $Q_3$ through other coefficients $Q$ given in Table \ref{coeff}. 

In the quantum limit of
large $\eta$ (sufficiently small $r_s$), $u_{\rm q} \approx Q_1 \eta$ 
which corresponds to a temperature-independent
contribution to $E_i$. Such a term strongly resembles the zero-point
energy of a harmonic Coulomb solid which has the same temperature and 
density dependence but a different coefficient. In the liquid,
the term in question is 
$(Q_1/\sqrt{3}) N_i \kB T_{\rm p} \approx 3.46 N_i \kB T_{\rm p}$ whereas
in the harmonic Coulomb solid 
$E_{\rm zero-point} = (3/2) u_1 N_i \kB T_{\rm p} \approx 0.767 N_i \kB T_{\rm p}$ 
where we have set $u_1 \approx 0.5114$ which is the first bcc phonon spectral moment (e.g., \citealt{BPY01}). We see 
that the coefficient in the liquid is 4.5 times larger than that in the 
bcc solid.

\begin{table*}
\caption[]{Fit coefficients in equation (\ref{fit_en}) $^{a)}$}
\renewcommand{\arraystretch}{1.4}
\begin{tabular}{ccccccccc}
\hline      
\hline      
    $A_1$ & $A_2$ & $B_1$ & $B_2$ & $B_3$ & $B_4$ & 
    $Q_1$ & $Q_2$ & $Q_4$  \\
\hline
   $-$0.9070 & 0.62954 &  0.00456 & 211.6 & $-$0.0001 & 0.00462 &
   5.994   & 70.3    & 22.7 \\
\hline      
\hline      
\end{tabular}
\\
\label{coeff}
$^{a)}$ Coefficients $A$ and $B$ are taken from \citet{PC00}; only coefficients $Q$ have been
varied to fit the data (see text for details).
\end{table*}

Formally, the fit (\ref{fit_en}) is based on the data at $\Gamma \leq \Gamma_{\rm m}$ 
($T\geq T_{\rm m}$), that is at temperatures higher than the melting temperature of 
a classic Coulomb liquid of ions. However, bearing in mind a sufficiently smooth dependence of $E_i$ on 
$T$ and $n_i$ we hope that we can extrapolate the fit to somewhat lower $T$ (larger $\Gamma$) 
to describe a supercooled Coulomb liquid or, actually, the ordinary quantum Coulomb liquid 
which in reality, due to quantum effects, solidifies at $\Gamma_{\rm m}$ 
slightly above the classic value of 175.

\subsection{Constructing thermodynamics}
\label{s:thermodyn}

Using the fit (\ref{fit_en}) we can construct analytic thermodynamics of quantum 
OCP of ions.

Multiplying equation (\ref{fit_en}) by $T$ and differentiating with respect 
to $T$ one obtains ionic isochoric heat capacity,
\begin{equation}
   \frac{C_{Vi}}{\kB N_i} \equiv  \frac{{\cal C}_{Vi}}{\kB n_i} = \frac{3}{2} + u_{\rm cl}(\Gamma)
   + u_{\rm q}(\eta) - \Gamma \frac{{\rm d}u_{\rm cl}}{{\rm d} \Gamma}
   - \eta \frac{{\rm d}u_{\rm q}}{{\rm d} \eta}~,
\label{CVgen}
\end{equation}
where ${\cal C}_{Vi}$ is the heat capacity per unit volume.
The Helmholtz free energy can be obtained by integration
\begin{equation}
    \frac{F_i}{N_i \kB T} = \frac{F_{\rm id}}{N_i \kB T} + 
    \int_0^\Gamma {\rm d} \Gamma' 
    \frac{u_{\rm cl}(\Gamma')}{\Gamma'}
    + \int_0^\eta {\rm d} \eta' 
    \frac{u_{\rm q}(\eta')}{\eta'}~,
\label{Fgen}
\end{equation}
where $F_{\rm id}$ is the Helmholtz free energy of an ideal Boltzmann 
gas of ions. Then the ion pressure can be found as     
\begin{equation}
    P_i = - \left( \frac{\partial F_i}{\partial V} \right)_T 
    = P_{\rm id} \left(  1 + \frac{1}{3} u_{\rm cl}
      + \frac{1}{2} u_{\rm q} \right)~,  
\label{Pgen}
\end{equation}
where $P_{\rm id} = n_i \kB T$ is the ideal gas pressure. Other practically important 
(see Section \ref{astro2}) 
quantities are the isochoric temperature derivative of pressure
\begin{equation}
  \left( \frac{\partial P_i}{\partial T} \right)_V = 
    n_i \kB \left( 1 + \frac{1}{3} u_{\rm cl} 
             + \frac{1}{2}  u_{\rm q}
    - \frac{\Gamma}{3} \frac{{\rm d}u_{\rm cl}}{{\rm d} \Gamma}
    - \frac{\eta}{2} \frac{{\rm d}u_{\rm q}}{{\rm d} \eta} \right)
\label{chigen}
\end{equation}
and the isothermal compressibility of ions
\begin{equation}
\left( \frac{\partial P_i}{\partial n_i} \right)_T = 
\kB T \left( 1 + \frac{1}{3} u_{\rm cl} 
+ \frac{1}{2}  u_{\rm q}
+ \frac{\Gamma}{9} \frac{{\rm d}u_{\rm cl}}{{\rm d} \Gamma}
+ \frac{\eta}{4} \frac{{\rm d}u_{\rm q}}{{\rm d} \eta} \right).
\label{dPidni}
\end{equation}

Evaluating integrals and derivatives in 
equations (\ref{CVgen}) -- (\ref{dPidni}) one obtains 
\begin{eqnarray}
    \frac{F_i}{N_i \kB T} &=& \frac{F_{\rm id}}{N_i \kB T} 
    + A_1 \sqrt{\Gamma (A_2+\Gamma)} 
    - A_1 A_2 \ln{\left(\sqrt{\frac{\Gamma}{A_2}}
    +  \sqrt{1+ \frac{\Gamma}{A_2}} \right)}  
\nonumber \\
      &+& 2 A_3 \left(\sqrt{\Gamma} - \arctan{\sqrt{\Gamma}} \right)
\nonumber \\
      &+& B_1 \Gamma - B_1 B_2 \ln{\left(1 + \frac{\Gamma}{B_2} \right)}
       + \frac{B_3}{2} \ln{\left(1 + \frac{\Gamma^2}{B_4} \right)} 
\nonumber \\
      &+& Q_1 \eta - Q_1 Q_2 \ln{\left(1 + \frac{\eta}{Q_2} \right)}
       + \frac{Q_3}{2} \ln{\left(1 + \frac{\eta^2}{Q_4} \right)}
\label{expl_F}
\end{eqnarray}
[note a typo in equation (16) of \citet{PC00} which should contain
$\Gamma/B_2$ rather than $\Gamma/B_1$ in one of the logarithms]. 
Furthermore,
\begin{eqnarray}
      \frac{C_{Vi}}{\kB N_i} &=& \frac{3}{2} 
      + \frac{\Gamma^{3/2}}{2} 
      \left[ \frac{A_3 (\Gamma-1)}{(1+\Gamma)^2} 
      - \frac{A_1 A_2}{(A_2+\Gamma)^{3/2}} \right]
\nonumber \\
      &+& \Gamma^2 
        \left[ \frac{B_3 (\Gamma^2 - B_4)}{(B_4 + \Gamma^2)^2} 
      - \frac{B_1 B_2}{(B_2 + \Gamma)^2} \right]
\nonumber \\
      &+& \eta^2 \left[ \frac{Q_3 (\eta^2 - Q_4)}{(Q_4 + \eta^2)^2} 
      - \frac{Q_1 Q_2}{(Q_2 + \eta)^2} \right]~,
\label{expl_CV}\\
    \frac{1}{\kB n_i} \frac{\partial P_i}{\partial T} &=& 1   
          + \frac{\Gamma^{3/2}}{6} 
      \left[ \frac{A_3 (\Gamma-1)}{(1+\Gamma)^2} 
      - \frac{A_1 A_2}{(A_2+\Gamma)^{3/2}} \right]
\nonumber \\
      &+& \frac{\Gamma^2}{3} 
       \left[ \frac{B_3 (\Gamma^2 - B_4)}{(B_4 + \Gamma^2)^2} 
      - \frac{B_1 B_2}{(B_2 + \Gamma)^2} \right]
\nonumber \\
      &+& \frac{\eta^2}{2} 
      \left[ \frac{Q_3 (\eta^2 - Q_4)}{(Q_4 + \eta^2)^2} 
      - \frac{Q_1 Q_2}{(Q_2 + \eta)^2} \right]~,
\label{expl_chi} \\
    \frac{1}{\kB T} \frac{\partial P_i}{\partial n_i} &=& 1   
+ \frac{\Gamma^{3/2}}{18} 
\left[ \frac{A_1 (9 A_2+8\Gamma)}{(A_2+\Gamma)^{3/2}} 
+ \frac{A_3(9+7\Gamma)}{(1+\Gamma)^{2}} \right]
\nonumber \\
      &+& \frac{\Gamma^2}{9} 
\left[ \frac{B_1 (5B_2+4\Gamma  )}{(B_2 + \Gamma)^2} 
+ \frac{B_3(5B_4+3\Gamma^2)}{(B_4 + \Gamma^2)^2} \right]
\nonumber \\
&+& u_q(\eta)-\frac{\eta^3}{4} 
\left[ \frac{Q_1 }{(Q_2 + \eta)^2} 
+ \frac{2Q_3\eta}{(Q_4 + \eta^2)^2} \right]~.
\label{expn_chi}
\end{eqnarray}
These fairly simple formulae give a complete first-principle 
description of the OCP thermodynamics at temperatures above the 
melting temperature neglecting ion statistics effects, which are 
expected to be very small under realistic conditions. For applications
to real systems, one has to add thermodynamic
quantities of the degenerate electron gas and  
contributions stemming from electron screening. The latter contributions are typically small and will be neglected here (e.g., \citealt{HPY2007}).

\section{Astrophysical implications}
\label{s:applic}

To illustrate astrophysical significance of new thermodynamics, let us 
describe accompanying modifications of physical properties of 
WD matter
which can result in potentially observable effects. We will 
consider liquid cores of WDs where the electrons are 
strongly degenerate. The degeneracy makes the electron gas almost
rigid and justifies the validity of the OCP model. By way 
of illustration, we consider a liquid of single species of 
fully ionized atomic nuclei. Actually, WD matter 
can contain a mix of different bare nuclei at high densities and ions in 
various ionization states at low densities but our present formalism 
does not allow us to study neither quantum effects in dense multispecies 
mixtures nor incompletely ionized plasma.

While analyzing WD  cores we will assume that they are
isothermal at a temperature $T$. This is a good approximation for
not too hot WDs because of high thermal conductivity of
degenerate electrons (e.g., \citealt{Schwarzschild,ST83}). 
When a WD cools, $T$ slowly decreases. For all realistic values 
of the internal temperature, the 
effective surface temperature $T_{\rm s}$ of a WD remains much 
lower than $T$ because of poor thermal conduction in its non-degenerate 
and mildly-degenerate outer envelope. The envelope is typically 
thin and low-massive; its thickness decreases with time. 

Warm WDs have liquid cores, the main subject of our
study. At a certain stage the core crystallyzes 
starting from the center. As the star cools, the crystallization front 
propagates to the stellar envelope.

\subsection{Heat capacity of liquid WD cores}
\label{astro1}
%

\begin{figure*}                                           
\begin{center}                                              
\leavevmode                                                 
\includegraphics[width=0.44\textwidth]{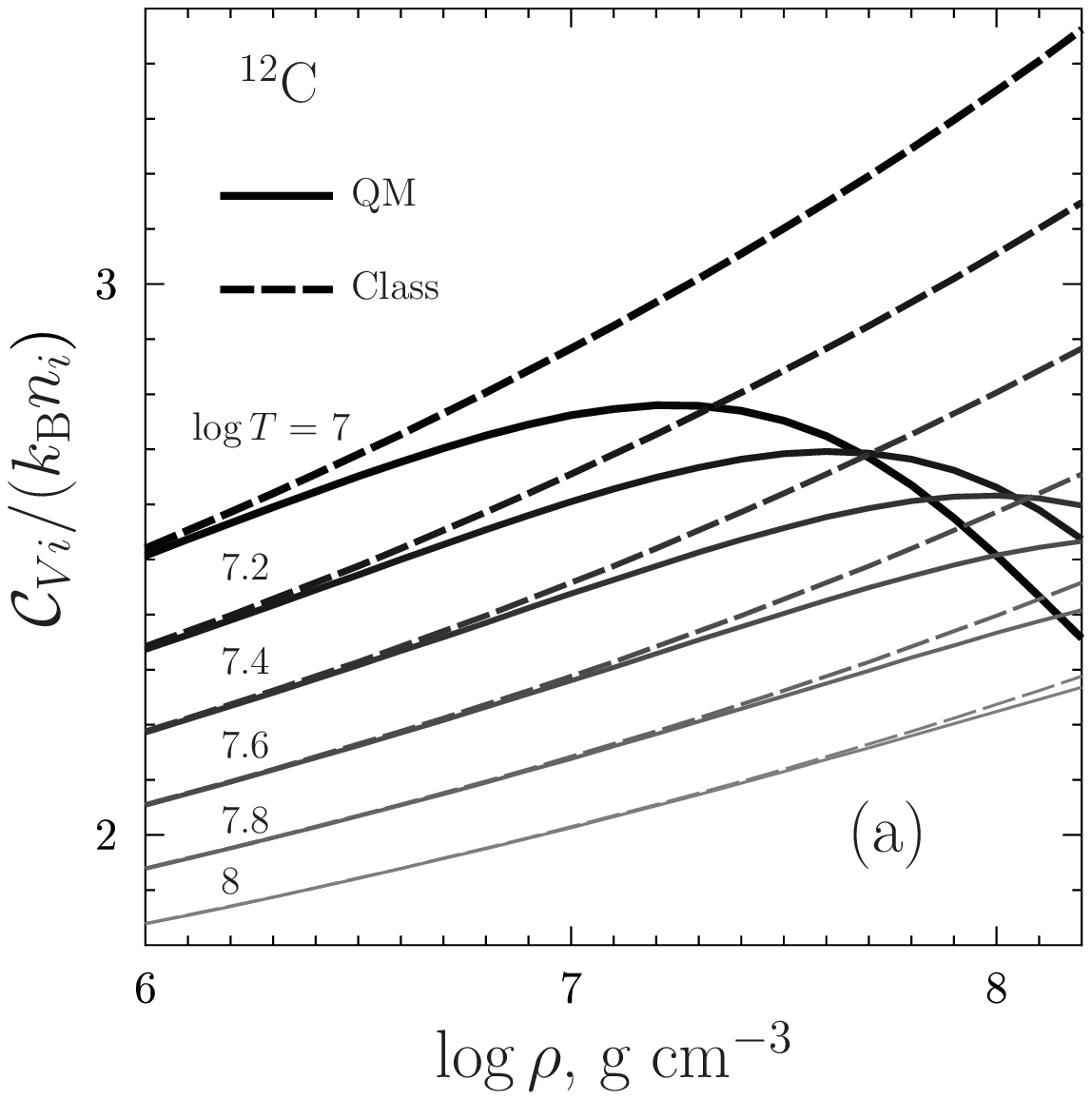}%
\hspace{5mm}
\includegraphics[width=0.44\textwidth]{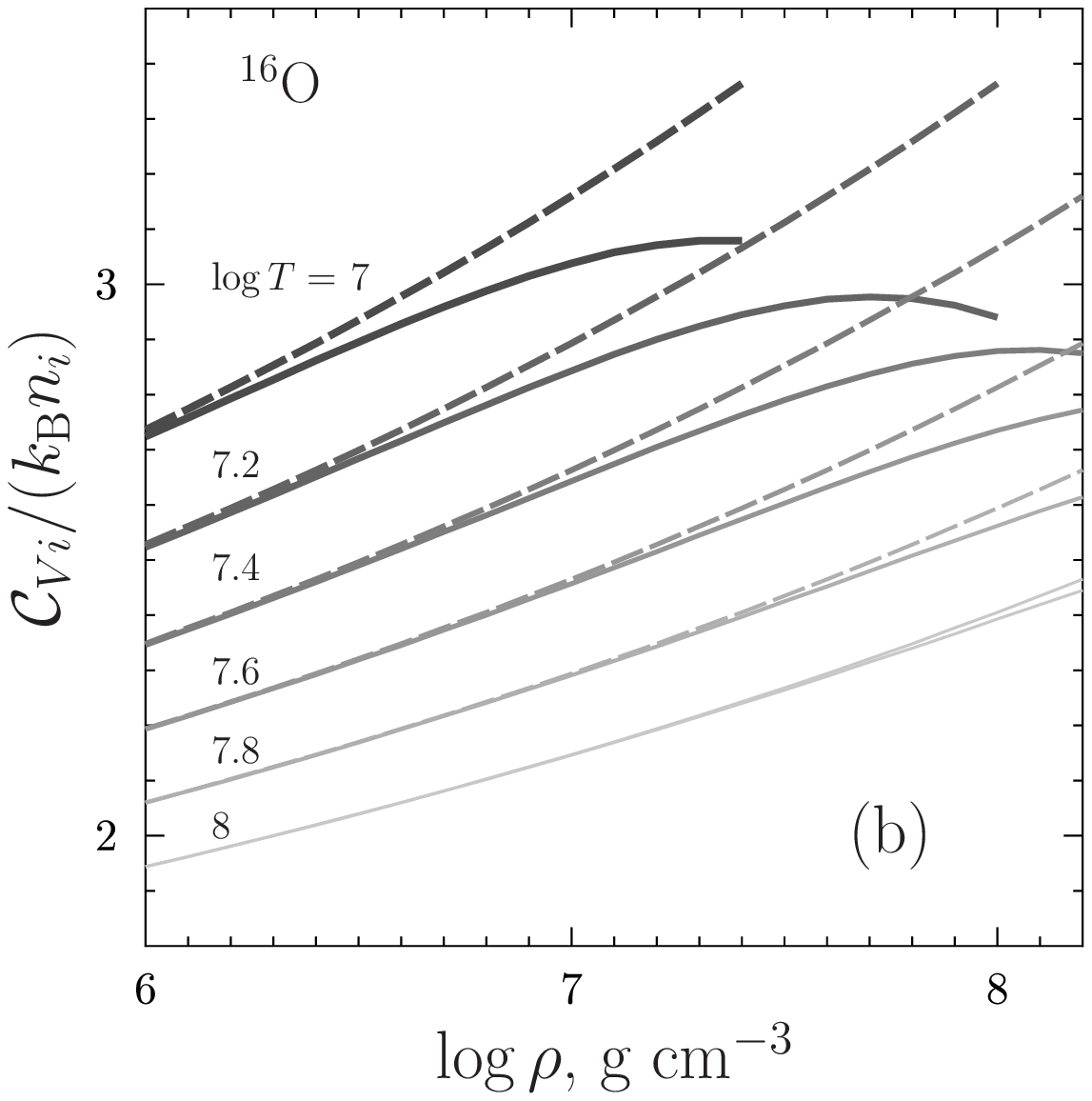}%
\vspace{2mm}
\includegraphics[width=0.44\textwidth]{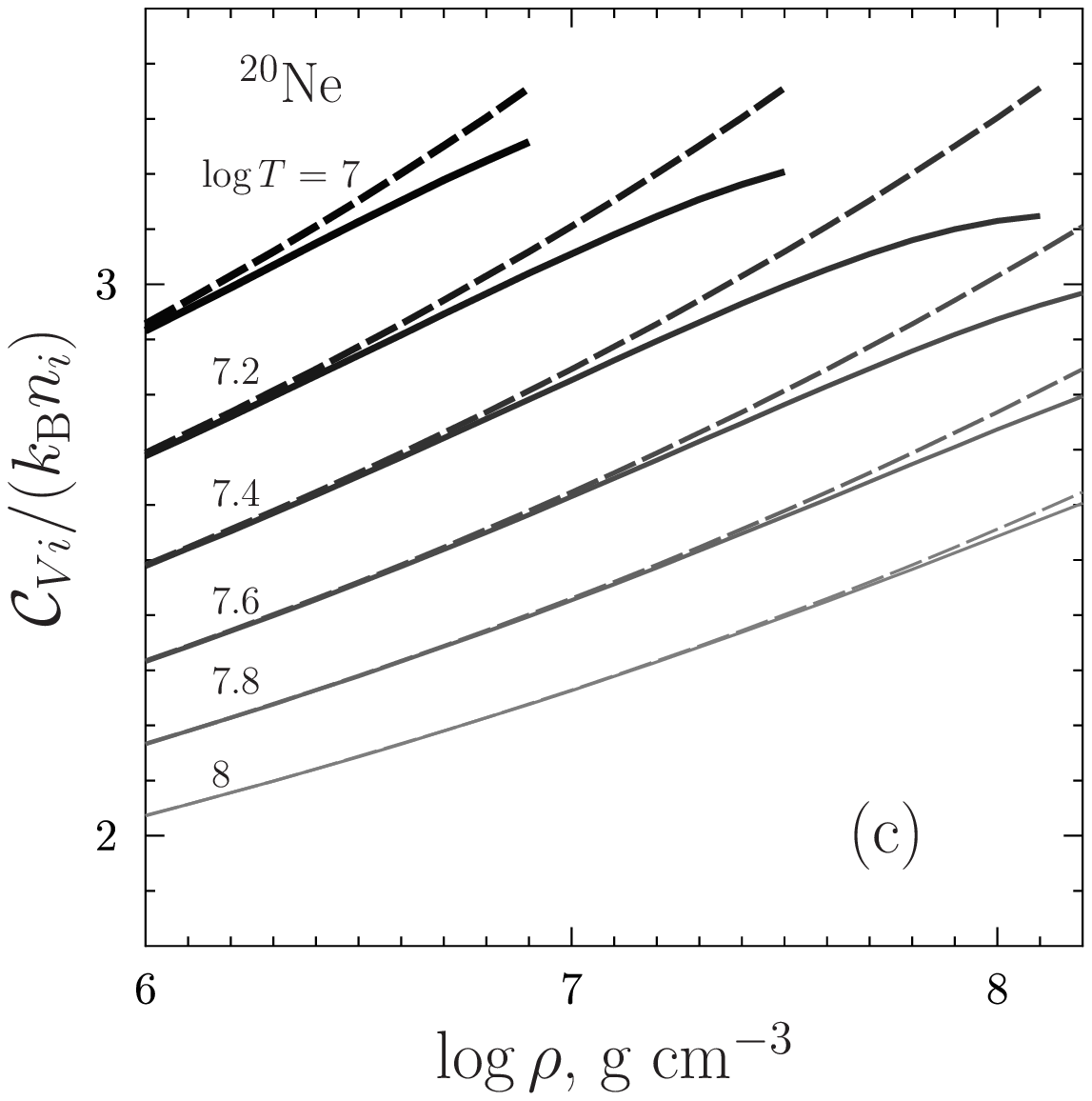}%
\hspace{5mm}
\includegraphics[width=0.44\textwidth]{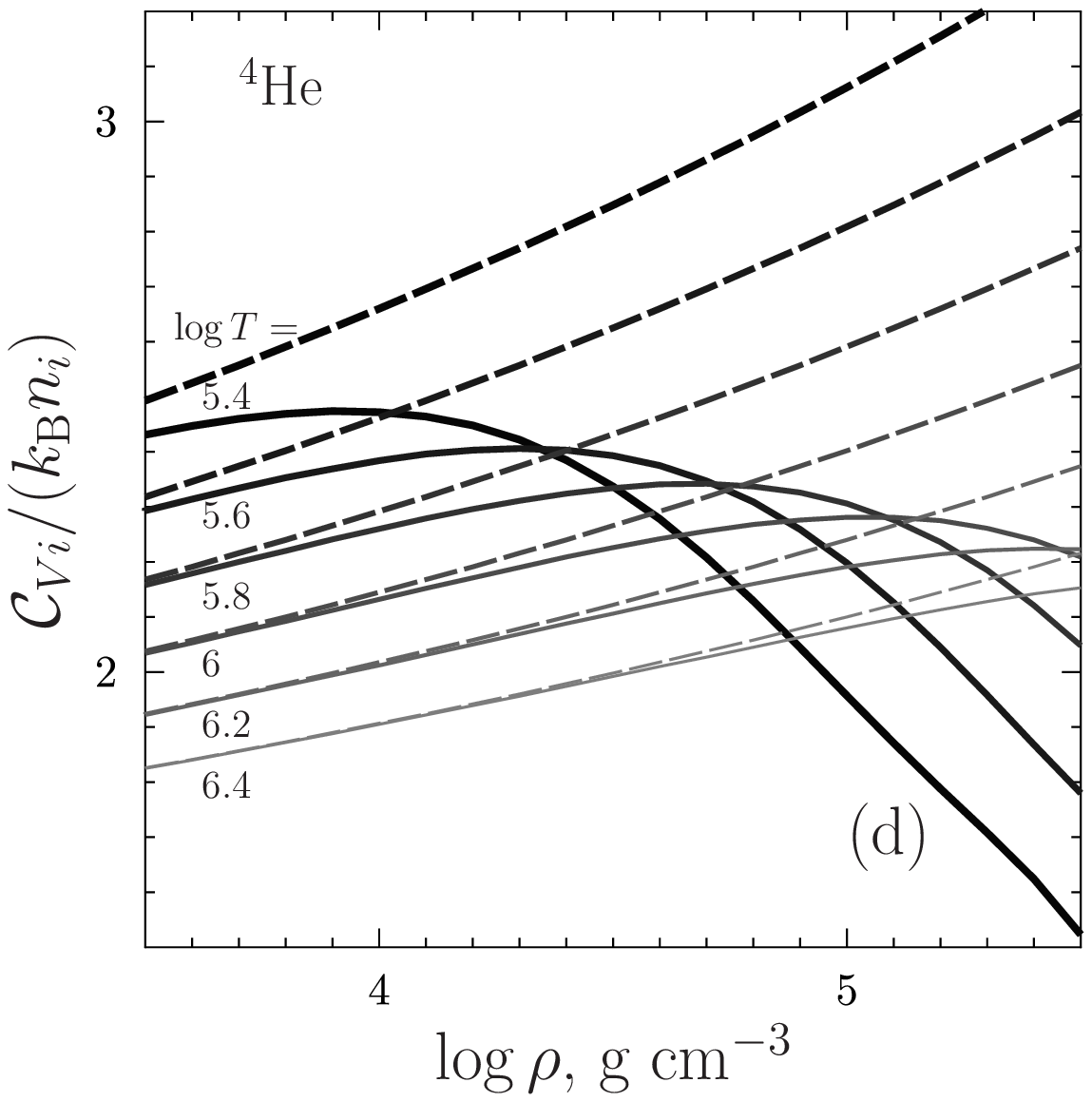}%
\end{center}                                                
\vspace{-0.4cm} 
\caption[]{Ionic heat capacity per one ion 
(in units of $\kB$) in a WD core composed 
of $^{12}$C (a), $^{16}$O (b), $^{20}$Ne (c) and $^{4}$He (d) as a 
function of mass density for several temperatures above 
crystallization; values of $\log T$ (in K) are indicated near the curves. 
Solid (dashed) curves include (disregard) quantum effects of ions.}                                             
\label{sh}
\end{figure*}
%

It is well known that ions in a WD core give 
the major contribution to the total heat capacity of the star.

In Fig.\ \ref{sh} we plot ionic heat capacity per one ion  
as a function of mass density $\rho$ for a range of internal 
WD temperatures $T$ above the 
melting temperature for four core compositions: $^{12}$C, $^{16}$O,
$^{20}$Ne and $^4$He  [panels (a), (b), (c) and (d), respectively]. 
Some low-temperature curves for oxygen and neon
are terminated at such densities where crystallization begins 
at a given $T$ ($T = T_{\rm m}$ for $\Gamma_{\rm m}=175$). 
Solid curves utilize full 
equation (\ref{expl_CV}), including quantum effects, 
while dashed curves represent classic 
heat capacity, i.e. the first two lines of equation (\ref{expl_CV}) only. 
As expected, quantum effects reduce the 
heat capacity; they are more pronounced for lighter elements, 
at higher densities 
and/or lower temperatures. 

\begin{figure*}                                           
\begin{center}                                              
\leavevmode                                                 
\includegraphics[width=0.44\textwidth]{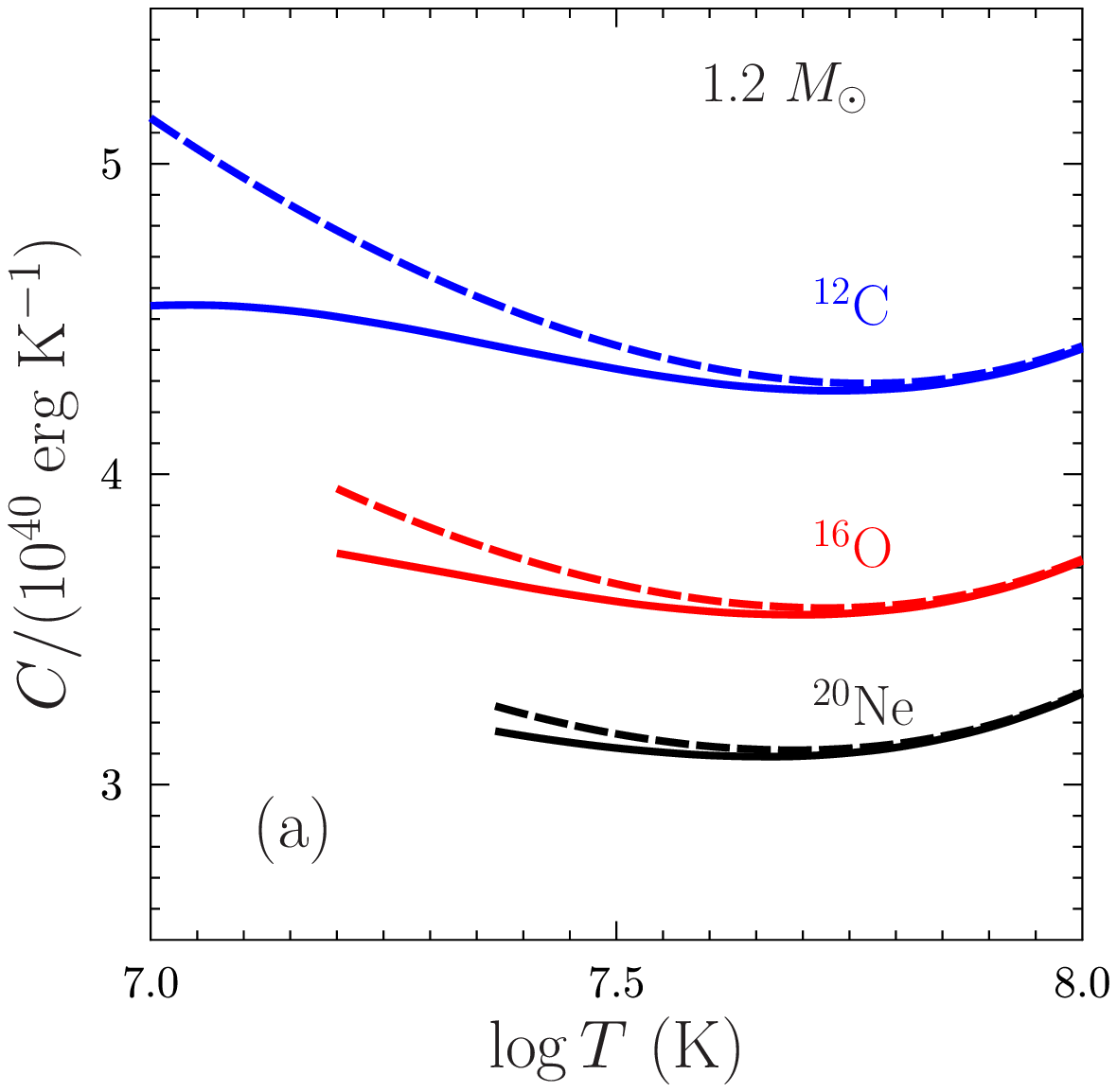}%
\hspace{5mm}
\includegraphics[width=0.44\textwidth]{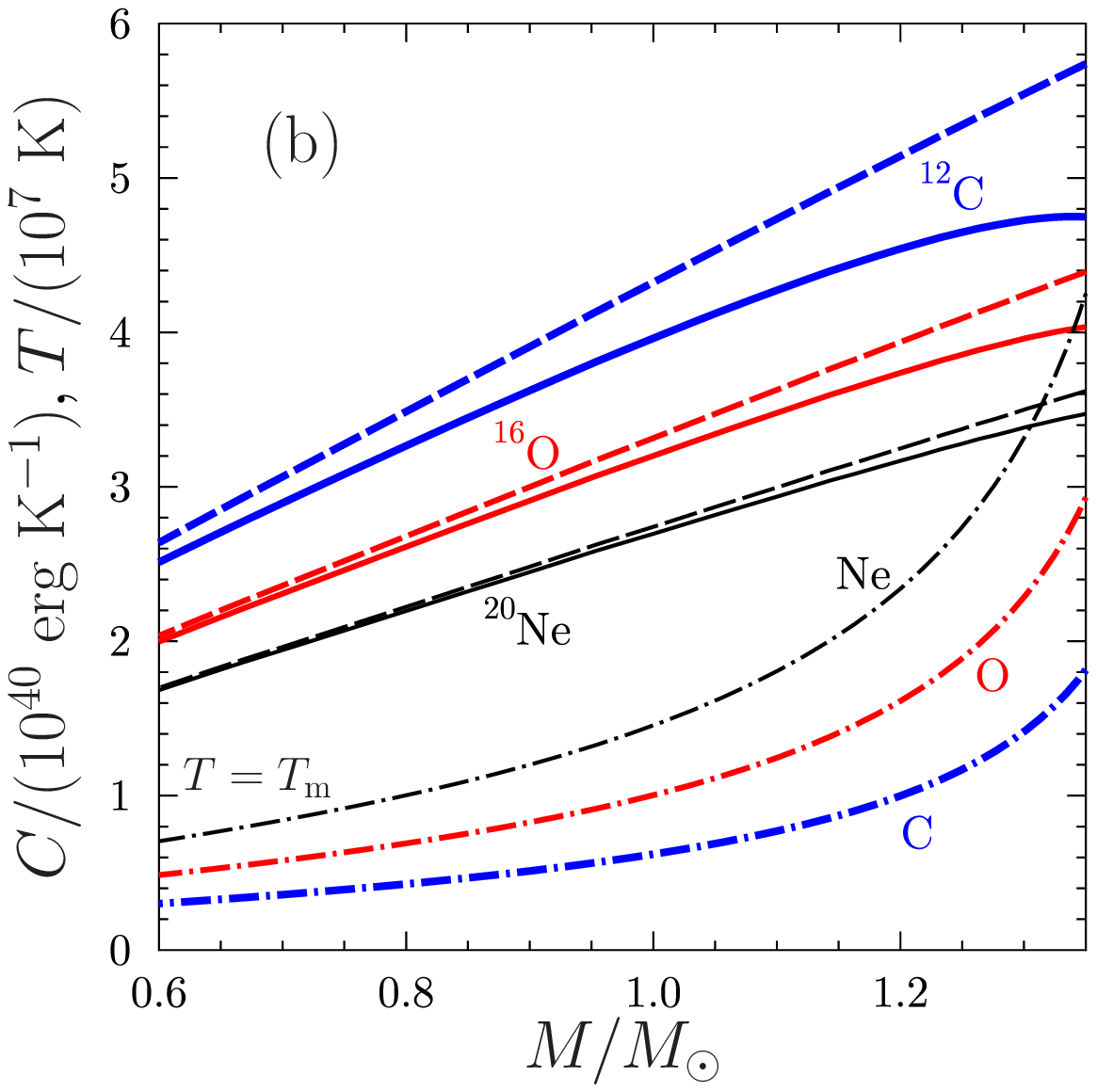}%
\vspace{2mm}
\includegraphics[width=0.44\textwidth]{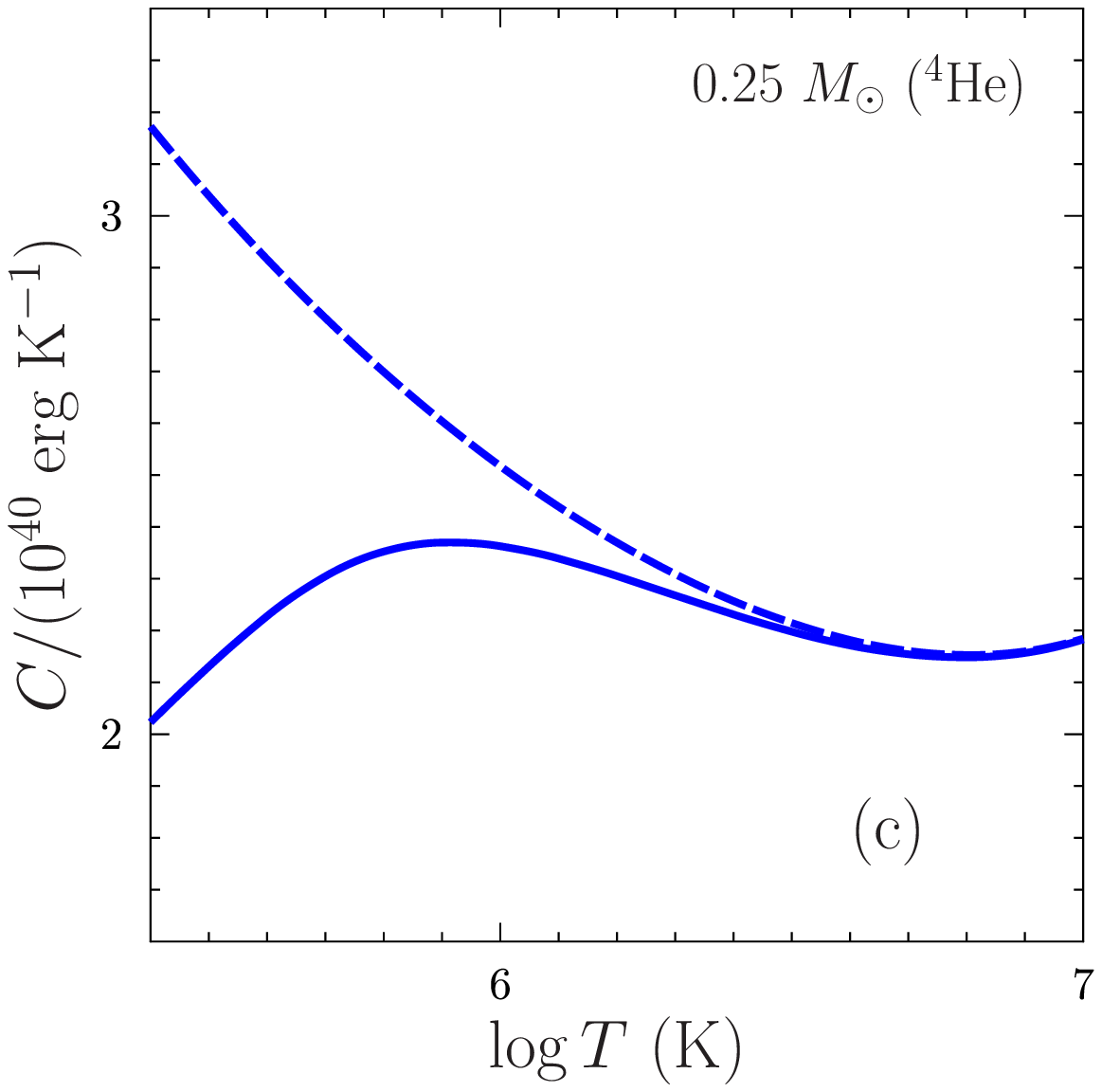}%
\hspace{5mm}
\includegraphics[width=0.44\textwidth]{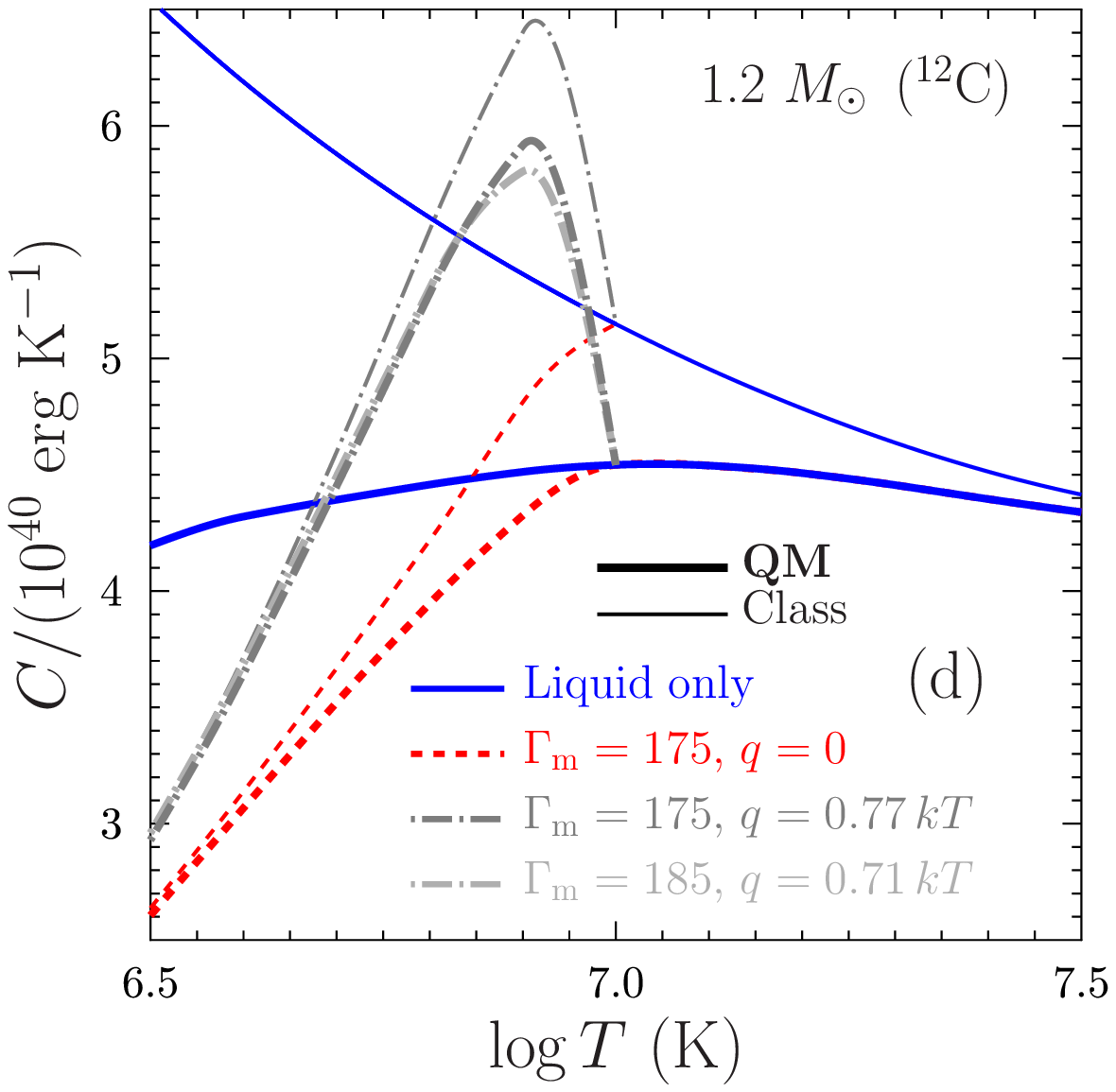}%
\end{center}                                                
\vspace{-0.4cm} 
\caption[]{(a) Total heat capacity of a 1.2 \Msun\ WD with 
isothermal liquid core of three compositions (marked near the curves) 
as a function of core temperature. Solid lines include and dashed lines 
exclude quantum effects of ions. (b) Total heat capacity of WDs
with different core compositions as a function of WD mass. At each mass, the
core temperature is equal to the melting temperature $T_{\rm m}$ in the WD center. Solid lines include and dashed lines 
exclude quantum effects. Three lower dash-dotted curves show respecive values of $T_{\rm m}$. (c) Same as in panel (a) but for a 
0.25 \Msun\ helium WD. (d) Effective heat capacity of a 
1.2 \Msun\ WD with isothermal carbon core as a function of 
core temperature with possible effects of crystallization and latent heat release 
(see text for details).}                                             
\label{hc}
\end{figure*}
%

Fig.\ \ref{hc} presents the total heat capacity $C(T)$ of liquid isothermal 
cores of WDs of different masses, core compositions and internal
temperatures,
\begin{equation}
     C(T)= \int {\rm d} V\, {\cal C}(\rho,T),
   \label{e:Ctot}
\end{equation}
where the integration is over the isothermal core, and ${\cal C}(\rho,T)$ is 
the heat capacity per unit volume; $C(T)$ is nearly equal to the total heat capacity
of the star (contribution of the outer envelope is negligible). 
Even though the ionic contribution is typically dominant, we have also included the
contribution of degenerate electrons, ${\cal C}={\cal C}_{Vi}+{\cal C}_{Ve}$. 
Because heat capacities of degenerate matter at constant volume and pressure 
are nearly the same (e.g., \citealt{HPY2007}), we can safely 
add the ion and electron contributions at constant volume.
The ion heat capacity per unit volume is calculated from equation (\ref{expl_CV}) as
${\cal C}_{Vi}=(C_{Vi}/N_i)n_i$. The electron contribution is
calculated with the aid of the standard Sommerfeld 
expansion \citep[e.g.,][]{LL80},
\begin{equation}
{\cal C}_{Ve} = 
\frac{\kB^2 {\rm \pi}^2 Tn_e}{p_{{\rm F}e} v_{{\rm F}e}}~,
\label{e:Ce}
\end{equation}
where $n_e$ is the electron number density, while $p_{{\rm F}e}$ 
and $v_{{\rm F}e}$ are the electron Fermi 
momentum and velocity, respectively. The envelope-core boundary 
is set at a density where the electron degeneracy temperature 
is three times higher than $T$. 

Since our figures are illustrative, our basic WD models 
are calculated using only the pressure of strongly degenerate electrons at $T=0$
(the only exception is Fig.\ \ref{f:dr}).
These models accurately reflect the main problem of WD study: at a given central 
density all WDs whose cores contain ions ($^4$He, $^{12}$C, $^{16}$O or $^{20}$Ne) 
with the same mass number to charge number
ratio ($A_i/Z_i=2$) have nearly identical internal mass density distributions and nearly
the same masses $M$ and radii $R$. This property complicates strongly the determination
of the internal composition of WDs from observations and motivates studies 
of such observational manifestations which are sensitive to the internal composition.

Integrated heat capacities $C(T)$ are good to this aim. They do depend on 
internal composition of WDs and they are important for WD 
cooling theory
(e.g., \citealt{Schwarzschild,ST83}). Because WD cooling is observable it can 
give valuable  information on the internal composition.  

Cooling of a WD with an isothermal core is governed by the equation
\begin{equation}
	\frac{{\rm d}T}{{\rm d}t}=-\frac{L}{C(T)},
\label{e:cool}
\end{equation}
where $L$ is the total thermal energy loss rate of the star which is a 
sum of the thermal surface luminosity $L_{\rm s}=4 \mathrm{\pi} \sigma R^2 T_{\rm s}^4$ 
($\sigma$ being the Stefan-Boltzmann constant) and the neutrino luminosity 
$L_\nu$ from the entire WD volume.  The isothermal-core approximation is 
usually valid in not too warm WDs (not in pre-WDs) where
neutrino cooling becomes inefficient and $L \approx L_{\rm s}$.

In Fig.\ \ref{hc}(a), we plot the integrated heat capacity of  
a 1.2\,\Msun\ ($R \approx 4180$ km) WD with a carbon, oxygen 
or neon core (blue, red or black curves, respectively) as a function of the 
core temperature (above melting). Solid (dashed) curves include 
(exclude) ion quantum effects. Once again one observes an 
amplification of the quantum effects with decreasing temperature 
and atomic number of ions. A general decreasing trend of the heat capacity
with increase of the atomic number is simply because $C(T)$ 
is proportional to the number of ions, and there is 
fewer heavier ions in a star of a fixed mass. Let us recall that
our figures are illustrative. Real WDs are 
thought to contain ionic mixtures. The lighter the 
ions, the larger $C(T)$ and the slower cooling for a fixed $M$ and
a fixed WD envelope model (which specifies $L_{\rm s}$).
This is well known. New result is that quantum effects in a massive
sufficiently cold WD substantially reduce $C(T)$ and
accelerate WD cooling.    

In Fig.\ \ref{hc}(b), we plot the total heat capacities of carbon, 
oxygen and neon WD cores as functions of stellar mass ranging 
from 0.6 to 1.35$M_\odot$. Again, solid curves include ion quantum effects
while dashed curves neglect them. 
For each mass, the temperature is set equal to the melting temperature in 
the WD center which maximizes the amplitude of the 
quantum effects attainable for a given mass and composition. The 
dot-dashed lines show these melting 
temperatures in units of $10^7$ K.     

In Fig.\ \ref{hc}(c), the same quantities as in panel (a) are plotted 
for a 0.25 \Msun\ ($R \approx 13450$ km) WD with the helium core. 
Clearly, in this case quantum effects can be very strong. In 
particular, as the temperature approaches crystallization in the WD  
center ($T_{\rm m} = 2.2 \times 10^5$ K at 
$\rho = 3.15 \times 10^5$ g cm$^{-3}$), 
the classic heat capacity overestimates the actual one by more than 
50 percent. Modifications of the WD heat capacity displayed in 
Fig.\ \ref{hc}(a)--(c) are expected to 
accelerate cooling of WDs, especially, the `most 
quantum' WDs with helium cores.

The above discussion assumed that crystallization of the ion
liquid took place at the classic value $\Gamma_{\rm m} = 175$.
However, quantum effects lead to a slow growth
of $\Gamma_{\rm m}$ with density \cite[e.g.,][]{JC96}. Besides, 
according to a preliminary report in Paper I, the latent heat $q$ released at 
crystallization decreases slowly with growing density, 
from 0.77$\kB T$ for classic OCP ($r_{\rm s} \to \infty$)
to 0.71$\kB T$ at $r_{\rm s} = 1200$ (which is smaller than in the center of a 1.2 \Msun\ carbon WD). In Fig.\ \ref{hc}(d) we aim 
to assess semi-quantitatively the effects of quantum modifications 
of $\Gamma_{\rm m}$ and $q$. To this end, we introduce an effective heat 
capacity of an isothermal core with temperature $T$, which 
incorporates the latent heat release,
\begin{equation}
    C_{\rm eff}(T) = C(T) - 4 \pi r^2 n_i q \, 
    \frac{{\rm d} r}{{\rm d} \rho} \, 
    \frac{{\rm d} \rho}{{\rm d} T_{\rm m}}~.
 \label{e:Ceff}
\end{equation}
In this case, radial coordinate $r$ as well as $n_i$, $q$, and $\rho$ 
must be taken at the crystallization front where $T_{\rm m}=T$. It is easy to show that
in the isothermal core approximation this effective heat capacity 
enters the WD cooling rate (\ref{e:cool}) automatically accounting for the latent heat release. This enables
one to compare relative importance of variations
of true heat capacity $C(T)$, latent heat, and melting temperature.

Fig.\ \ref{hc}(d) demonstrates the temperature dependence of the effective heat capacity of a 1.2 \Msun\ WD with carbon core, calculated under different assumptions. 
All thick lines 
are computed including quantum effects in the ion liquid, while thin lines are calculated without these effects. The solid (blue) lines 
are obtained by completely disregarding crystallization 
(assuming supercooled liquid at $T \leq T_{\rm m}$). As expected, 
quantum effects progressively reduce the heat capacity with lowering $T$. 
The short-dashed (red) lines are calculated 
using the harmonic-lattice model for the heat capacity of 
bcc crystals \citep[][]{BPY01} (at those $\rho$ where $T_{\rm m}>T$ for $\Gamma_{\rm m}=175$) but neglecting the latent heat ($q=0$). 
The heat capacity of crystalline ions   
is strongly suppressed by quantum effects which is known to accelerate cooling of 
old and cold WDs (e.g., \citealt{ST83}).
Finally, the dot-dashed (gray) lines show the same as the short-dashed ones but taking into account the latent heat release. Dark gray lines assume
classic crystallization temperature ($\Gamma_{\rm m}=175$)
and latent heat ($q=0.77\,\kB T_{\rm m}$).
The light gray thick dot-dashed 
curve corresponds to $\Gamma_{\rm m}=185$ and $q=0.71 \kB T$ in the 
entire star. It represents an overestimation of quantum 
modifications of the melting parameters because $r_{\rm s}=1200$ is 
never reached in a 1.2 \Msun\ carbon WD. Even then, the quantum 
modifications of the melting parameters do not appear significant.

The latent heat release at crystallization is known to delay cooling of 
sufficiently old WDs, prior to a cooling acceleration 
in a quantum crystal. The effect was predicted by \citet{LV1975} and
is supported \citep{TFF19} by the DR2 {\it Gaia} data on old massive DA WDs. 
According to  Fig.\ \ref{hc}(d), quantum suppression of the
heat capacity of WDs with liquid carbon cores, which accelerates cooling prior to the crystallization, 
reduces the effect of the latent heat 
(which delays cooling). If, however, the central part of 
the WD core contains not carbon, but heavier elements such
as $^{16}$O or/and $^{20}$Ne, quantum suppression of the 
heat capacity would be weaker [cf.\ Fig.\ \ref{hc}(b)]. In any case the suppression seems to
be not strong enough to completely eliminate the latent heat effect.

\subsection{Compressibilities and WD seismology}
\label{astro2}

Besides cooling evolution, many WDs demonstrate rich spectra
of pulsations (e.g., \citealt{WK08,CABK19}). Characteristic pulsation
periods are about some minutes. They are generally thought to be 
non-radial $g$-modes with multipolarity $\ell$ from 1 to about 5
and with the number of radial pulsation nodes $k$ from 1 to about 50. 
These pulsations are generated in WD
envelopes, but they can penetrate deeply into degenerate cores. Their studies
(comparison of observations and theory) provide useful information 
about WD parameters (masses, radii,
rotation) and internal composition and allow one to test basic
principles of fundamental physics, such as possible variations
of fundamental physical constants with time.
 
In order to see the impact of new microphysics on pulsational properties of WDs it seems natural to analyze its effect on the
\BV\ frequency ${\cal N}$, the basic quantity for seismological studies which
is a local characteristic of stellar matter expressed through
thermodynamic quantities. According 
to \citet{BFWKT91} the 
\BV\ frequency can be cast in the following form
\begin{equation}
    {\cal N}^2 = \frac{g^2 \rho}{P} \frac{\chi_T}{\chi_\rho}
         (\nabla_{\rm ad} - \nabla)~,
\label{BVdef}
\end{equation}
where $g$ is a local gravity at a density $\rho$; $P$ is the pressure; $\chi_T$ and
$\chi_\rho$ are compressibilities of matter, and $\nabla_{\rm ad}$ is the
adiabatic gradient. Finally, $\nabla$ is the actual local logarithmic pressure derivative of 
the temperature. The quantities $\chi_T$, $\chi_\rho$ and
$\nabla_\mathrm{ad}$ are thermodynamic and can be expressed as (e.g., \citealt{HPY2007})
\begin{eqnarray}
      \chi_T &=& \left(\frac{\partial \ln{P}}{\partial \ln{T}} 
      \right)_\rho~,
\label{e:chiT}\\
      \chi_\rho &=& \left(\frac{\partial \ln{P}}{\partial \ln{\rho}} 
      \right)_T~,
\label{e:chirho}\\
      \nabla_{\rm ad} &=& \left(\frac{\partial \ln{T}}{\partial \ln{P}} 
      \right)_S=\frac{\chi_T}{\chi_T^2+\chi_\rho {\cal C}_V/P}~,
\label{e:Dad}      
\end{eqnarray}
where $S$ is the entropy. In a WD core containing OCP of ions we write $P=P_e+P_i$ and ${\cal C}_V={\cal C}_{Ve}+{\cal C}_{Vi}$.
In the approximation of the isothermal core, $\nabla \approx 0$. 
The overwhelming contribution to the pressure is due to 
degenerate electrons at $T=0$ \citep[e.g.,][]{LL80}. Adding also 
first-order thermal correction given by the Sommerfeld expansion we have 
\begin{equation}
     P_e = P_0 \left[ \sqrt{1+x^2} \left(\frac{2 x^3}{3}  - x \right)
      +\ln{\left(x + \sqrt{1+x^2} \right)} + \frac{4}{9}
      {\rm \pi}^2 \left( \frac{T}{T_0} \right)^2 \frac{x(x^2+2)}{\sqrt{1+x^2}} \right]~,
\label{Pe}
\end{equation}
where $P_0 = m_e^4 c^5/(8 \pi^2 \hbar^3) = 1.801 \times 10^{23}$ dyn 
cm$^{-2}$, $T_0=m_ec^2/\kB \approx 5.930 \times 10^9$~K and $x = p_{{\rm F}e}/(m_e c)$ is the  
relativity parameter of strongly degenerate electrons [$m_e$ is the electron mass;
$p_{{\rm F}e}=\hbar (3 {\rm \pi}^2 n_e)^{1/3}$]. 

Now we can easily compute $\chi_T$, $\chi_\rho$, $\nabla_{\rm ad}$ and ${\cal N}$.
The electron and ion pressures are given by equations (\ref{Pe}) and (\ref{Pgen}).
The respective temperature derivative needed for $\chi_T$ is 
\begin{equation}
    \left( \frac{\partial P_e}{\partial T} \right)_V = 
    \frac{8 \pi^2 P_0 T x (2+x^2)}{9 T_0^2 \sqrt{1+x^2}}
\end{equation}
for electrons and is given explicitly in 
equation (\ref{expl_chi}) for ions. The density derivatives needed
for $\chi_\rho$ are given by
\begin{equation}
  \frac{\partial P_e}{\partial \rho}= \frac{8 P_0 x^5}{9 \rho \sqrt{1+x^2}}
  \left[ 1 + \frac{{\rm \pi}^2 (2+3 x^2 + 2 x^4)}{6 x^5 (1+x^2)} 
  \left( \frac{T}{T_0} \right)^2  \right] 
\end{equation}
for electrons (because, with high accuracy, $\rho=M_i n_i=M_i n_e/Z_i$) 
whereas ${\partial P_i}/{\partial \rho}=({\partial P_i}/{\partial n_i})/M_i$ 
where ${\partial P_i}/{\partial n_i}$ can be taken from (\ref{expn_chi}). The heat capacities 
per unit volume ${\cal C}_{Ve}$ and ${\cal C}_{Vi}$ needed in $\nabla_{\rm ad}$ are given by
equations (\ref{e:Ce}) and (\ref{expl_CV}), respectively.

\begin{figure*}                                           
\begin{center}                                              
\leavevmode                                                 
\includegraphics[width=0.45\textwidth]{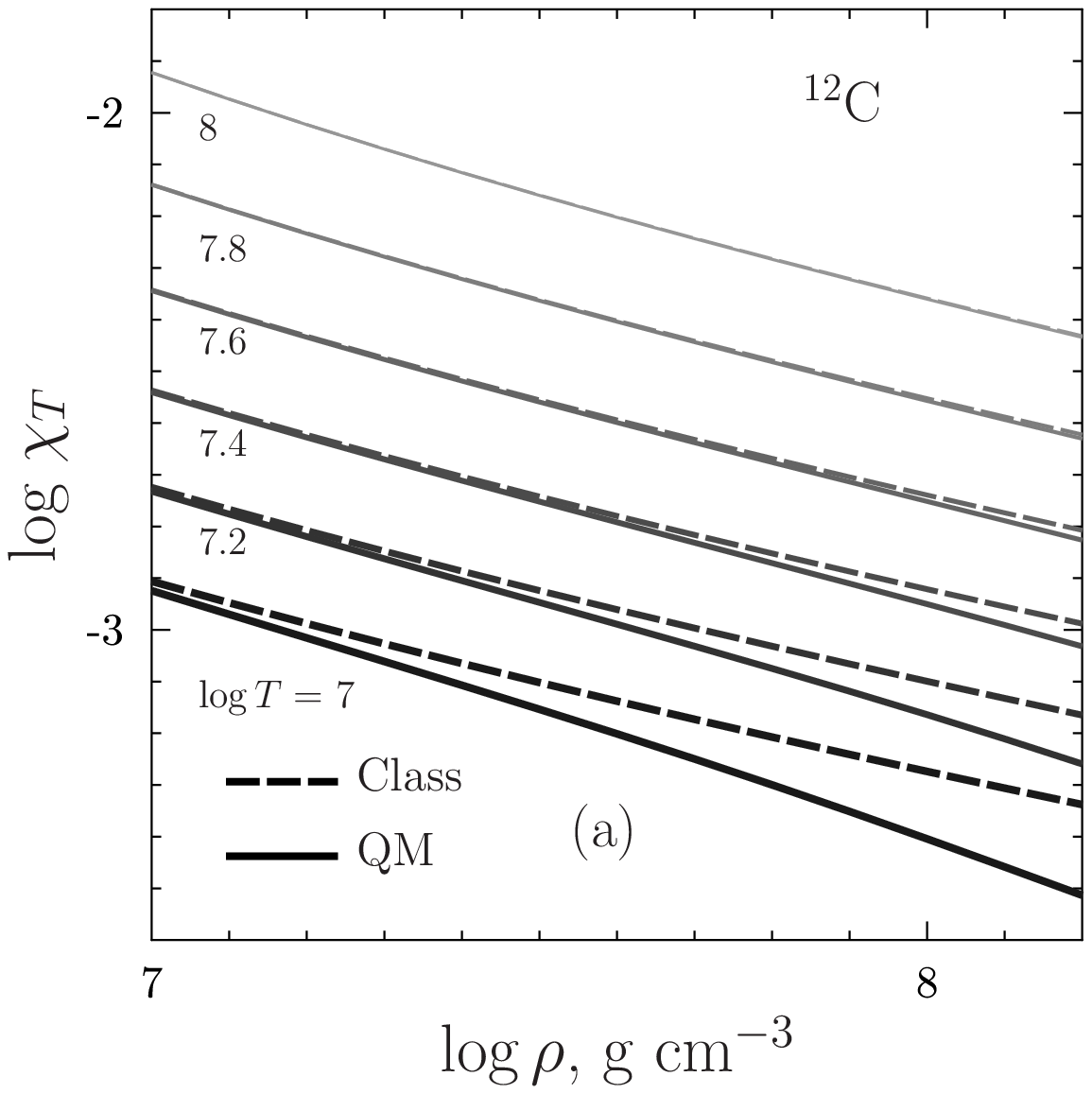}%
\hspace{5mm}
\includegraphics[width=0.45\textwidth]{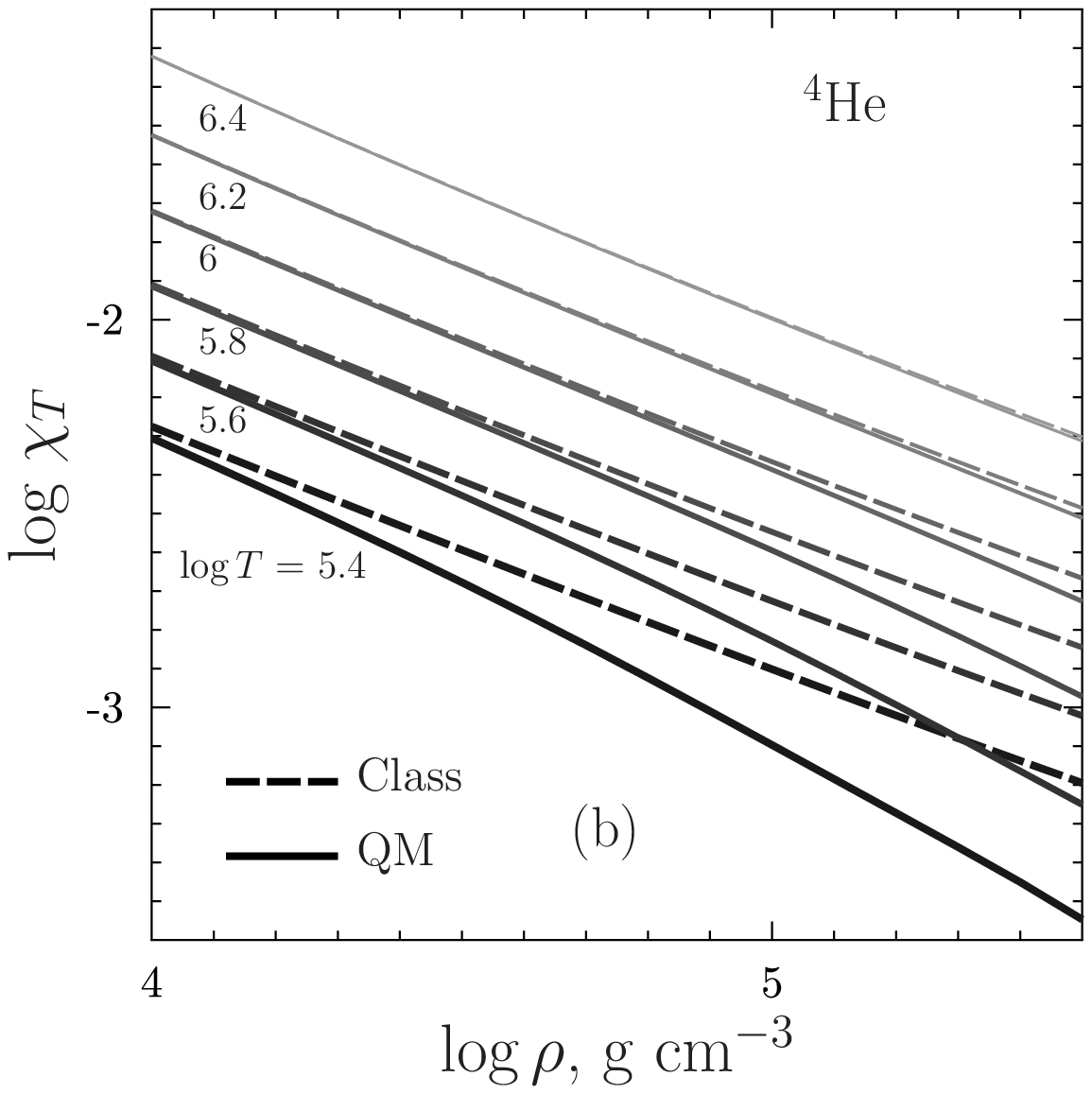}%
\end{center}                                                
\vspace{-0.4cm} 
\caption[]{Logarithm of compressibility $\chi_T$ of isothermal WD  
matter composed of $^{12}$C (a) or $^4$He (b) as a function of $\log \rho$
for several values of temperature. Solid (dashed) curves 
include (exclude) quantum effects in the Coulomb liquid of ions.}                                             
\label{xit}
\end{figure*}
%

The compressibilities $\chi_\rho$ and $\chi_T$ are interesting not only for WD
seismology. In particular they determine the adiabatic gradient $\nabla_{\rm ad}$ which
regulates convective stability of stellar matter. Evidently, in degenerate WD cores
 $\chi_\rho$ is mainly determined by the bulk pressure of degenerate electrons;
it is affected by the state of ions but only weakly. On the contrary,
$\chi_T$ is mainly determined by ions. It is plotted in 
Fig.\ \ref{xit} as a function of mass density for carbon (a) and 
helium (b) matter at several temperatures above melting. Note that the scale 
of the vertical axis is logarithmic. Solid lines are calculated using full 
equation (\ref{expl_chi}) while dashed lines are respective classic values.
Quantum effects are seen to decrease $\chi_T$.  
As in the case of the heat capacity, the strongest decrease
occurs in deeper WD layers and at lower temperatures.

\begin{figure}                                           
\begin{center}                                              
\includegraphics[width=78mm,clip]{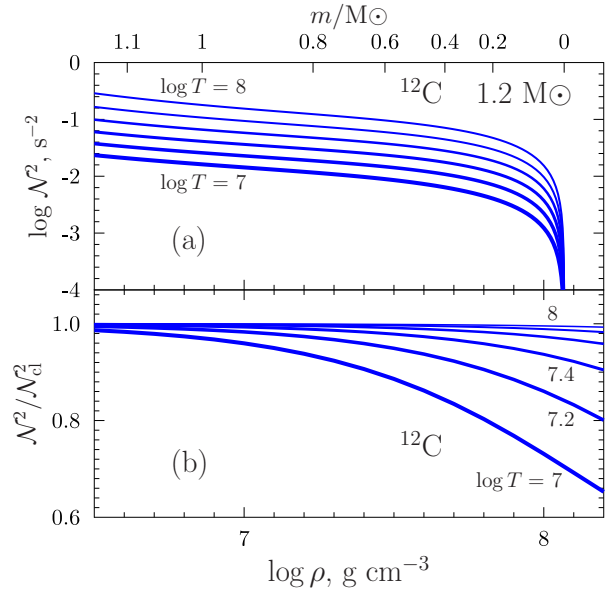} 
\end{center}                                                
\caption[]{(a) Logarithm of the squared \BV\ 
frequency in the carbon core of an isothermal 1.2 \Msun\ WD as a function of mass density for six values of the core 
temperature, from $T=10^7$ to $10^8$ K. The upper horizontal
scale displays accumulated stellar mass $m$. (b) The ratio of squared 
\BV\ frequencies calculated including and neglecting quantum
effects of ions versus $\rho$  
at the same temperatures.}                                             
\label{BVpic}
\end{figure}
%

Fig.\ \ref{BVpic}(a) shows the density dependence of squared \BV\ frequency 
in the central part of an isothermal carbon core of a
1.2 \Msun\ WD as a function of density $\rho$
at six selected temperatures, with $\log T$ [K]=7, 7.2, \ldots 8. 
The upper horizontal axis displays values of stellar mass $m$ (in units of \Msun)
accumulated in spheres restricted by respective $\rho$. 
To demonstrate the importance of quantum effects of ions, 
Fig.\ \ref{BVpic}(b) shows the ratio of the squared 
\BV\ frequency 
with quantum effects to that without quantum effects as a  
function of mass density at the same $T$. Since gravity 
cancels out, this becomes a universal function of density, 
temperature, and composition and does not depend on a particular WD model.
Ion quantum effects lower ${\cal N}$ and actual $g$-mode frequencies.

\begin{figure}                                           
\begin{center}                                              
\leavevmode                                                 
\includegraphics[width=78mm,clip]{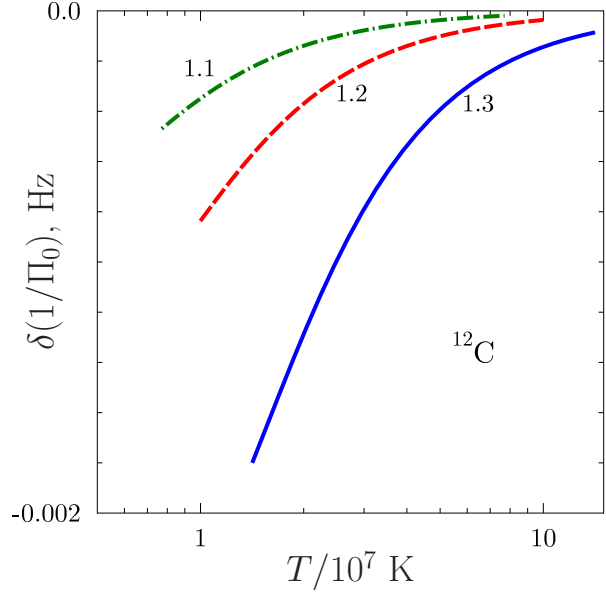} 
\end{center}                                                
\vspace{-0.4cm} 
\caption[]{Variation of $\Pi_0^{-1}$ that determines frequency 
difference between two $g$-modes with large successive 
numbers of radial nodes ($k$, $k+1\gg 1$) under the
effect of quantum effects of ions  as a function of core 
temperature for carbon WDs with $M=$1.1, 1.2 and 1.3 \Msun. 
At low $T$ the curves are broken at onset of crystallization in the WD center.}                                              
\label{dPi0}
\end{figure}
%

In order to gauge the effect of these differences on the pulsational 
properties of WDs we consider the asymptotic period spacing
\citep[e.g.,][]{GCAWC19} between pulsation modes with
fixed multipolarity $\ell$ and azimuthal number $m$ but 
succesive numbers $k$ and $k+1$ of radial nodes in the
limit of large $k$,  
\begin{eqnarray}
  \Delta \Pi_{l}^{\rm a} &=& \Pi_0/\sqrt{\ell(\ell+1)}~,
\label{Pi1} \\
      \Pi_0 &=& 2 {\rm \pi}^2 \left[\int_0^R {\rm d}r \,
      \frac{\cal N}{r} \right]^{-1}~.
\label{Pi0}
\end{eqnarray}
According to Fig.\ \ref{BVpic}(b), the difference between ${\cal N}$ and 
${\cal N}_{\rm cl}$ vanishes  
while still at fairly high densities. This allows us to calculate 
the quantity
\begin{equation}
   \delta \left( \frac{1}{\Pi_0} \right) = \frac{1}{2\pi^2} 
   \int_0^R {\rm d}r \, \frac{({\cal N} - {\cal N}_{\rm cl})}{r},
\end{equation}
which is insensitive to thermodynamics in the 
WD envelope and outer core. This quantity determines 
the magnitude of eigenmode frequency 
change originating from the quantum phenomena in the ion liquid
near the WD center. 
In Fig.\ \ref{dPi0}, $\delta (\Pi_0^{-1})$ is plotted as a function of 
core temperature (in units of $10^7$ K, above melting) for carbon WDs of masses 1.1, 1.2, and 1.3 \Msun.
According to numerical data of \citet{GCAWC19}, the range of realistic 
values of $\Delta \Pi_1^{\rm a}$ can be roughly estimated as 
$\sim 5$--50 s. Per equation (\ref{Pi1}) with $\ell=1$, this translates into $\Pi_0^{-1} \sim 14$--140 mHz
which should be compared with the expected variation displayed in Fig.\ \ref{dPi0}. 
Given the remarkable accuracy of seismological measurements, an 
experimental verification of spectral variations due to quantum effects
predicted in Fig.\ \ref{dPi0} seems feasible.

\subsection{Radii of low-mass WDs with He cores}
\label{astro3}

\begin{figure}                                           
	\begin{center}                                              
	\leavevmode                                                 
	\includegraphics[height=78mm,clip]{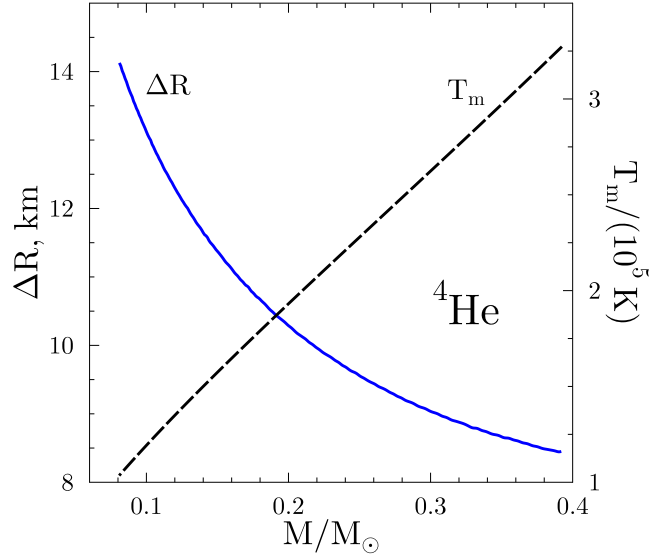} 
	\end{center}                                                
	\vspace{-0.4cm} 
	\caption[]{Radius increase $\Delta R$ (solid line) due to quantum ion effects for
    WDs with $^4$He cores versus WD mass. In each
	case the temperature of the isothermal core corresponds to the crystallization in the WD center $T=T_{\rm m}$. 
  This temperature is plotted by the dashed line along the right vertical axis.}                                              
	\label{f:dr}
\end{figure}
%

Here we comment on the importance of ion quantum effects for hydrostatic models of 
low-mass WDs. To this aim, we have calculated a sequence of WD models   
of different masses $M$ with isothermal helium cores, from
0.1 \Msun\ to 0.4 \Msun. In doing so we
have taken into account the ion pressure at a finite core temperature
$T$. This temperature has been set equal
to the crystallyzation temperature in the WD 
center (the lowest temperature at which the entire core
remains liquid). This temperature 
increases with the growth of $M$ and is shown by a dashed curve in 
Fig.\ \ref{f:dr} referring to the right vertical axis. 
For each $M$, we have calculated
WD models twice: including quantum term in the ion pressure (\ref{Pgen}) 
and assuming purely classic ions.

Fig.\ \ref{f:dr} presents the difference of WD radii $\Delta R$,
calculated with and without ion quantum effects versus
mass $M$. The ion pressure is much smaller than the electron 
one and the quantum contribution to the ion pressure is, in turn, smaller 
than the absolute magnitude of the ion electrostatic contribution. 
Nevertheless, the difference due to quantum effects is seen to be 
$\sim 10$ km. Somewhat counterintuitively, it increases with decreasing
mass, i.e. $\Delta R$ is larger for more classic extremely low-mass 
helium stars! The decrease is associated with two effects, a decrease 
of $|P_i|/P_e$ when $\rho$ increases in heavier stars and a general 
increase of $R$ with decrease of $M$. 

The effect is small but may be important for AM Canum Venaticorum 
variables (AM CVn) which constitute a class of binaries with extremely 
short orbital periods (down to a few minutes); e.g., see
\citet{Deloye07,S10,Ramsay18}. Their evolution is governed by mass transfer 
from a donor star to an accreting WD as well as by orbital momentum loss 
due to gravitational wave emission. We are interested in the scenario 
where the donor is a low-mass ($M \lesssim 0.4$ \Msun) He dwarf. It 
can be very cold ($T_{\rm s}$ as low as $\sim 1000$ K, 
see \citealt{Deloye07}). In the process of accretion, it loses mass and
moves to the left in Fig.\ \ref{f:dr} so that the predicted $\Delta R$
due to quantum effects increases. Thus accurate incorporation of these
effects may be crucial to correctly model
the final outcome of the evolution, in particular, when or whether 
there will be a direct impact and merger.

\section{Conclusion}
\label{s:conclude}

In Section \ref{s:fit}, we present a simple analytical formula, equation (\ref{fit_en}), which 
accurately approximates calculations of the energy 
of a quantum OCP of ions in a uniform 
electron background. The calcuations were performed in 
Paper I from first principles in the regime of a strongly-coupled liquid 
($1\leq \Gamma \leq 175$) neglecting the effects of quantum
statistics of ions. At $\Gamma < 1$, equation (\ref{fit_en}) reduces to 
the familiar expression for a classic liquid \citep[][]{PC00}
combined with the first Wigner-Kirkwood correction. In this way
equation (\ref{fit_en}) stays valid at small $\Gamma$. 

Equation (\ref{fit_en}) allows one to construct accurate analytic thermodynamics
of ionic OCP at $\Gamma\leq 175$, obtain
an analytic expression for the Helmholtz 
free energy, equation (\ref{expl_F}), and calculate 
temperature and density derivatives which are required for finding 
various thermodynamic 
quatities for astrophysical applications.  

In Section \ref{s:applic}, we apply the results for constructing 
major thermodynamic quantities in an OCP of ions immersed in a degenerate
gas of electrons (including Sommerfeld temperature corrections for electron
thermodynamic quantities). This is a good model to describe dense matter
in degenerate cores of WDs with account of quantum effects
in ion motion. We apply this formalism to investigate the ionic quantum effects 
on the total heat capacity of WDs, compressibilities $\chi_T$ and $\chi_\rho$,
\BV\ frequency of WDs,
and radii of low-mass WDs with helium cores. Our main conclusion is
that quantum effects of ions become really important in central regions
of WD cores at low enough temperatures (within a factor of a few of the crystallization temperature).
In general, the strongest effects occur in WDs with massive helium or carbon
cores. In particular, quantum effects can noticeably decrease the
heat capacity and accelerate cooling of massive WDs with carbon cores 
prior to crystallization.
The only exception to this general trend is an increase of the quantum correction 
to the radius of a helium WD, $\Delta R$, with decrease of its mass (Fig.\ \ref{f:dr}). 

The theory we present is valid for quantum OCP of ions neglecting 
ion exchange effects which are not relevant for applications. Further work is required to elaborate
this theory by studying quantum anharmonic corrections to the harmonic-lattice
model of a crystallized phase and by developing detailed theory
of quantum ion mixtures. 
 
\section*{Acknowledgments}
We are grateful to L. R. Yungelson who drew our attention to the problems of AM CVn binaries.
 
\bibliographystyle{mnras} 

\end{document}